\documentclass[twocolumn,showpacs,prl,nofootinbib,superscriptaddress,preprintnumbers,amsmath,amssymb]{revtex4}

\usepackage{graphicx}
\usepackage{dcolumn}
\usepackage{bm}

\newcommand{\gray}{$\gamma$-ray}

\newcommand{\Xco}{$X_{\rm CO}$}

\newcommand{\gardian}{{\it GaRDiAn}{}}
\newcommand{\fermi}{{\it Fermi}{}}

\begin{document}

\preprint{}

\title{\fermi{} Large Area Telescope Measurements of the Diffuse Gamma-Ray Emission at Intermediate Galactic Latitudes}

\author{A.~A.~Abdo}
\affiliation{Space Science Division, Naval Research Laboratory, Washington, DC 20375, USA}
\affiliation{National Research Council Research Associate, National Academy of Sciences, Washington, DC 20001, USA}
\author{M.~Ackermann}
\affiliation{W. W. Hansen Experimental Physics Laboratory, Kavli Institute for Particle Astrophysics and Cosmology, Department of Physics and SLAC National Accelerator Laboratory, Stanford University, Stanford, CA 94305, USA}
\author{M.~Ajello}
\affiliation{W. W. Hansen Experimental Physics Laboratory, Kavli Institute for Particle Astrophysics and Cosmology, Department of Physics and SLAC National Accelerator Laboratory, Stanford University, Stanford, CA 94305, USA}
\author{B.~Anderson}
\affiliation{Santa Cruz Institute for Particle Physics, Department of Physics and Department of Astronomy and Astrophysics, University of California at Santa Cruz, Santa Cruz, CA 95064, USA}
\author{W.~B.~Atwood}
\affiliation{Santa Cruz Institute for Particle Physics, Department of Physics and Department of Astronomy and Astrophysics, University of California at Santa Cruz, Santa Cruz, CA 95064, USA}
\author{M.~Axelsson}
\affiliation{Department of Astronomy, Stockholm University, SE-106 91 Stockholm, Sweden}
\affiliation{The Oskar Klein Centre for Cosmo Particle Physics, AlbaNova, SE-106 91 Stockholm, Sweden}
\author{L.~Baldini}
\affiliation{Istituto Nazionale di Fisica Nucleare, Sezione di Pisa, I-56127 Pisa, Italy}
\author{J.~Ballet}
\affiliation{Laboratoire AIM, CEA-IRFU/CNRS/Universit\'e Paris Diderot, Service d'Astrophysique, CEA Saclay, 91191 Gif sur Yvette, France}
\author{G.~Barbiellini}
\affiliation{Istituto Nazionale di Fisica Nucleare, Sezione di Trieste, I-34127 Trieste, Italy}
\affiliation{Dipartimento di Fisica, Universit\`a di Trieste, I-34127 Trieste, Italy}
\author{D.~Bastieri}
\affiliation{Istituto Nazionale di Fisica Nucleare, Sezione di Padova, I-35131 Padova, Italy}
\affiliation{Dipartimento di Fisica ``G. Galilei", Universit\`a di Padova, I-35131 Padova, Italy}
% ...

\collaboration{The \fermi{} LAT Collaboration}
\noaffiliation

\date{\today}


\begin{abstract}
The diffuse Galactic \gray{} emission is produced by cosmic rays (CRs)
interacting
with the interstellar gas and radiation field.
Measurements by the Energetic Gamma-Ray Experiment Telescope (EGRET) 
instrument on the {\em Compton Gamma-Ray Observatory} indicated
excess \gray{} emission $\gtrsim 1$ GeV relative to 
diffuse Galactic \gray{} emission models 
consistent with directly measured CR spectra 
(the so-called ``EGRET GeV excess''). 
The excess emission was observed in all directions on the sky,
and a variety of explanations have been proposed,
including beyond-the-Standard-Model scenarios like annihilating or decaying 
dark matter.
The Large Area Telescope (LAT) instrument on the \fermi{} 
Gamma-ray Space 
Telescope has measured the diffuse 
\gray{} emission with improved 
sensitivity and resolution compared to EGRET.
We report on LAT measurements of the diffuse \gray{} 
emission for energies 100 MeV to 10 GeV and Galactic latitudes
$10^\circ \leq |b| \leq 20^\circ$.
The LAT spectrum for this region of the sky is well reproduced by a 
diffuse Galactic \gray{} emission model that is 
consistent with local CR spectra
and inconsistent with the EGRET GeV excess.

\end{abstract}

\maketitle


{\it Introduction:} 
The diffuse \gray{} emission, both Galactic and extragalactic, is of
significant interest for astrophysics, particle physics, and cosmology.
The diffuse Galactic emission (DGE) is produced by interactions of 
cosmic rays (CRs), mainly protons and electrons, 
with the interstellar gas (via $\pi^0$-production and bremsstrahlung) and 
radiation field (via inverse Compton [IC] scattering) \cite{Ginzburg1964,1971NASSP.249.....S}.
It is a direct probe of CR fluxes in distant locations, and
may contain signatures of physics beyond the Standard Model, 
such as dark matter annihilation or decay.
The DGE is a foreground for point-source detection and hence influences the 
determination of the source positions and fluxes.
It is also a foreground for the much fainter extragalactic
component, which is the sum of contributions from unresolved sources
and truly diffuse
emission, including any signatures of large scale structure formation,
emission produced by 
ultra-high-energy CRs interacting with
relic photons, and many other processes (e.g.,~\cite{Dermer2007} and references
therein).
Therefore, understanding the DGE is a necessary first step in all such studies.

The 
excess diffuse emission $\gtrsim$ 1 GeV
in the Energetic Gamma-Ray Experiment Telescope (EGRET) 
data \citep{Hunter1997} relative to that expected from
DGE models consistent with the directly measured CR nucleon and 
electron spectra \citep{Hunter1997,Strong2000} led to the 
proposal that this emission was the 
long-awaited signature of dark matter annihilation~\citep{deBoer2005}.
More conventional interpretations included variations of CR
spectra in the Galaxy~\citep{Hunter1997,Porter1997,Strong2000}, 
contributions by unresolved point sources
\citep{BV2004}, and instrumental 
effects~\citep{Hunter1997,Moskalenko2007,Stecker2008}.

A model of the DGE depends on the CR spectra throughout the 
Galaxy as well as the distribution of the target gas and interstellar 
radiation field (ISRF).
Starting from the distribution of CR sources and particle injection spectra,
the distribution of CRs throughout the Galaxy is determined taking into 
account relevant energy losses and
gains, then the CR distributions are folded with the target
distributions to calculate the 
DGE \citep[e.g.,][]{SMR04}.  
Defining the inputs and calculating the models are not trivial
tasks and involve analysis of data from
a broad range of astronomical and astroparticle
instruments~\citep{StrongAnnRev2007}.

The \fermi{} Large Area Telescope (LAT) was launched on June 11, 2008.
It is over an order of magnitude more sensitive than its 
predecessor, EGRET, with a more stable response due to the lack of 
consumables.
The LAT data permit more detailed studies of the DGE than have been 
possible ever before. 

In this paper, analysis and results for the DGE are shown for
the Galactic mid-latitude range $10^\circ \leq |b| \leq 20^\circ$ 
measured by the LAT in the first 5 months of the science phase of the mission.
This region was chosen for initial study since it maximises the fraction
of signal from DGE produced within several kpc of the Sun and 
hence uncertainties associated with CR propagation, knowledge
of the gas distribution, etc., should be minimised.
The calculation of the DGE at lower Galactic latitudes requires CR fluxes
throughout the whole Galaxy and thus is model dependent, 
while the
emission at higher latitudes is more affected by contamination from 
charged particles misclassified as photons and uncertainties in the model 
used to estimate the DGE.
The diffuse emission at lower and higher Galactic latitudes 
will be addressed in subsequent LAT papers.

{\it LAT Data Selection and Analysis:}
The LAT is a pair-conversion telescope with a precision tracker and 
calorimeter, each consisting of a $4\times 4$ array of 16 modules, a segmented 
anti-coincidence detector (ACD) that covers the tracker array, 
and a programmable 
trigger and data acquisition system. 
Full details of the instrument, onboard and ground data processing, and other 
mission-oriented support are given in \cite{InstrumentPaper}.

The 
data selection used in this paper is made using the standard 
LAT ground processing and background rejection scheme \cite{InstrumentPaper}.
This consists of two basic parts: first a simple accept-or-reject selection 
(prefiltering) followed by a classification tree (CT) \cite{1984CTbook} 
based determination of the relative probability of being background or 
signal. 
The prefiltering phase screens particles entering the LAT
for their charge neutrality using the tracker and ACD.
The direction reconstruction software extrapolates particle trajectories
found in the tracker back to the scintillation tiles of the ACD, and we
accept only events in which the intersected tiles show no significant signal.
In addition, the prefiltering phase includes considerations of the shape of the 
calorimeter shower energy deposition and how well the found tracks project
into the energy centroid.
The overall background rejection of the prefiltering phase is $10^3 - 10^4$
depending on energy, 
yielding an 
efficiency $> 90\%$ 
for \gray{s} that convert into 
electron-positron pairs in the LAT.

Classification trees, which afford an efficient and 
statistically robust method for distinguishing signal from noise, are used to 
reduce backgrounds further.
Using quantitites defined from ACD, tracker, and calorimeter data, 
the CTs are trained on Monte Carlo simulated data which have passed the 
prefilter 
described above.
Multiple CTs are built to make the procedure robust 
against statistical fluctuations during the training procedure. 
The result from averaging the output from these CTs is the probability for 
an event to
be a photon or background.
This final selection parameter allows the signal purity to be 
set according to the needs of the analysis.
For the analysis of diffuse emission, the cut on the CT generated 
probability is set 
such that the Monte Carlo prediction of the orbit-averaged 
background rate is $\sim0.1$ Hz integrated over the full instrument 
acceptance $> 100$ MeV.
This yields a \gray{} efficiency $> 80\%$, and the residual background 
is at a level where the majority of the contamination arises from irreducible
sources such as \gray{s} produced 
by CR interactions in the passive material outside the ACD, e.g., 
the thermal blanket and micrometeroid shield of the 
LAT (see Fig.~13 in \cite{InstrumentPaper}).
The events corresponding to the above criterion are termed ``Diffuse'' class and
are the standard low-background event selection.

The analysis presented here uses post-launch instrument response functions
(IRFs).
These take into account pile-up and accidental coincidence 
effects in the detector 
subsystems that are not considered in the definition of the pre-launch IRFs. 
Cosmic rays, primarily protons,
pass through the LAT at a high rate and sufficiently 
near coincidences with \gray{s}
leave residual signals that can result in \gray{s} being misclassified,
particularly at energies $\lesssim 300$ MeV.  
The post-launch IRFs were
derived using LAT events read from a special trigger that 
produces periodic detector readouts, irrespective of the 
signals present, as a background overlay on
the standard simulations of \gray{s} and provide an accurate
accounting for the instrumental pile-up and accidental coincidence effects.  
The on-axis effective area for the event selection used in this paper is 
$\sim 7000$ cm$^2$ at 1 GeV and is energy dependent; this is approximately 10\%
lower at 1 GeV than the pre-launch effective area corresponding to the 
same event selection.
The systematic uncertainties of the effective area, evaluated
by comparing the efficiencies of analysis cuts for data and simulation of
observations of Vela, are also energy dependent: 10\% below 100 MeV, 
decreasing to
5\% at 560 MeV, and increasing to 20\% at 10 GeV and above.  
The point spread function (PSF) and energy resolution are as described in 
\cite{InstrumentPaper}.

The LAT nominally operates in a scanning mode that covers the whole sky 
every two orbits (i.e., ~3 hrs). 
We use data taken in this mode from the commencement of scientific 
operations in mid-August 2008 to the end of December 2008. 
The data were prepared using the LAT Science Tools 
package, 
which is available from the \fermi{} Science Support 
Center \cite{FSSC}. 
Events satisfying the Diffuse class selection and coming from 
zenith angles $< 105^\circ$ (to greatly reduce the contribution by 
Earth albedo
\gray{s}) were used. 
To further reduce the effect of Earth albedo 
backgrounds, the time intervals when the Earth was appreciably within the 
field of view (specifically, when the centre of the field of view was 
more than 47$^\circ$ from the zenith) were excluded from this analysis.
This leaves 9.83 Ms of total livetime in the data set.
The energy-dependent exposure was calculated using the IRFs described above.

The photon counts and exposure were further processed using the \gardian\ 
package,
part of a suite of tools we have developed to analyse the DGE where 
the analysis 
approach is described in \cite{gadget_and_gardian}, with
more details to be given in a subsequent publication.
Gamma-ray skymaps were generated using a 
HEALPix \cite{healpix} 
scheme at order 7 (i.e., $\sim0.2^\circ$ resolution) with 
5 bins per decade in energy from 100 MeV to 10 GeV. 
For each energy bin the intensity was obtained by dividing the in-bin counts by 
the spectrally-weighted exposure over the bin.
We used two methods for the spectral weighting: a power law with index $-2$ and
the spectral shape of the assumed DGE model (described below).
With the energy binning used in this paper the differences in the derived 
intensities were $< 1$\% between these two weighting schemes.

\begin{figure}[ht]
\includegraphics[width=8.5cm]{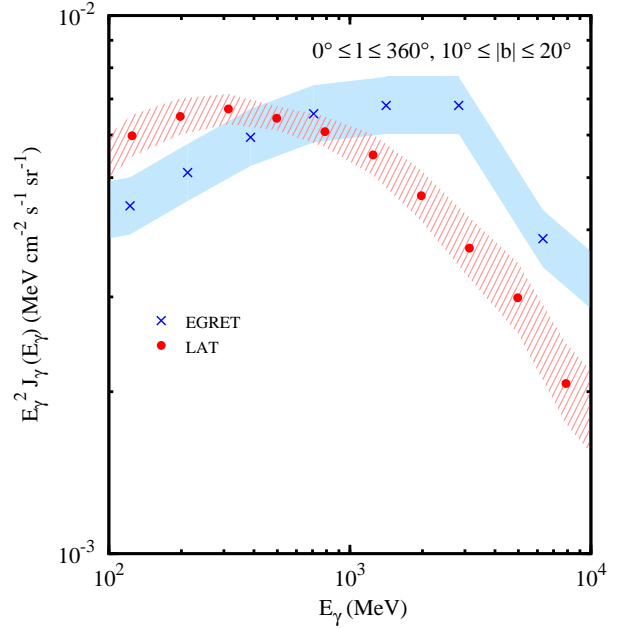}
\caption{\label{fig:data}Diffuse emission intensity averaged over all Galactic
longitudes for latitude range $10^\circ \leq |b| \leq 20^\circ$. 
Data points: LAT, red dots; EGRET, blue crosses.
Systematic uncertainties: LAT, red; EGRET, blue.}
\end{figure}

Figure~\ref{fig:data} shows 
the LAT data averaged over all Galactic longitudes
and the latitude range $10^\circ \leq |b| \leq 20^\circ$.
The hatched band surrounding the LAT data indicates the systematic 
uncertainty in the measurement due to the uncertainty in the effective area 
described above.
Also shown are the EGRET data for the same region 
of sky derived from count maps and exposures 
available via the {\em CGRO} Science Support Center \cite{CSSC} 
and processed 
following the procedure described in \cite{SMR04} and we have included
the standard systematic uncertainty of 13\% \cite{Esposito1999}.
For both data sets the contribution by point sources has not been subtracted.
The LAT-measured spectrum is significantly softer than the EGRET measurement 
with an integrated intensity $J_{\rm LAT}(\geq 1 \, {\rm GeV}) = 
2.35\pm0.01\times10^{-6}$ cm$^{-2}$ s$^{-1}$ sr$^{-1}$ 
compared to the EGRET integrated intensity 
$J_{\rm EGRET}(\geq 1\, {\rm GeV}) = 3.16\pm0.05\times10^{-6}$ 
cm$^{-2}$ s$^{-1}$ sr$^{-1}$ where the errors are statistical only.
Not included in the figure is the systematic uncertainty in the energy scale, 
which is conservatively estimated from comparison between Monte Carlo and 
beam test data as $< 5$\% for 100 MeV to 1 GeV, and $< 7$\% above 1 GeV 
where it is believed that if any bias is present energies are overestimated.
Taking the uncertainty on the energy scale into account, the LAT spectrum could
be softer, increasing the discrepancy with the EGRET spectrum further.

\begin{figure}[ht]
\includegraphics[width=8.5cm]{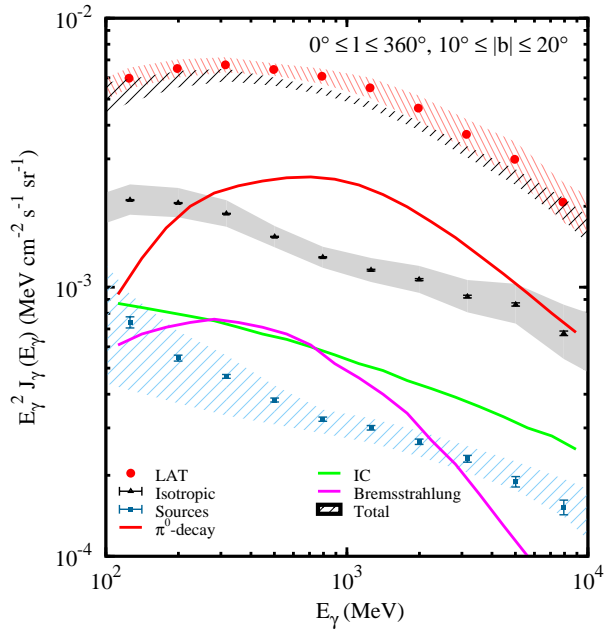}
\caption{\label{fig:model}LAT data with model, source, and 
UIB components for sky region in Fig.~\ref{fig:data}.
Model (lines): $\pi^0$-decay, red; bremsstrahlung, magenta; IC, green.
Shaded/hatched regions: UIB, grey/solid; source, blue/hatched; 
total (model + UIB + source), black/hatched.}
\end{figure}

Figure~\ref{fig:model} compares the LAT spectrum shown in Fig.~\ref{fig:data} 
with the 
spectra of an {\em a priori} DGE model, and a point-source 
contribution and unidentified background (UIB) component 
derived from fitting the LAT data that are described
below.
The DGE model is an updated version of the ``conventional'' 
model from GALPROP \cite{SMR04}. 
Major
improvements include use of the formalism and corresponding code for pion 
production in $pp$-interactions by
\cite{2006ApJ...647..692K}, 
a complete recalculation of the ISRF \cite{Porter2008}, updated gas maps,
and an improved line-of-sight integration routine. 
However, it is still an {\em a priori} model that is based on local
cosmic-ray data, and does not use \gray{} data.
Table~\ref{tab:table1} summarises 
the numerical values by energy bin for
the different components shown in Fig.~\ref{fig:model}.

The source and UIB components were obtained by fitting 
the LAT data using \gardian\ with the DGE model held constant. 
Point source locations were taken from the 3 month \fermi{} LAT source list
down to 
sources with 5-$\sigma$ significance.
Due to the limited statistics of all but the very brightest sources, we used 
3 bins per energy decade in the fitting procedure. 
Source positions were fixed but the spectra were fit using one free 
parameter for the source flux per energy bin.
The UIB component was determined by fitting the data
and sources over all Galactic longitudes for the high-latitude region 
$|b| \ge 30^\circ$ for the full LAT energy range shown in the figure.
Using this high-latitude region minimises the effect of contamination by the 
bright Galactic ridge which can be significant even up to $\sim 10^\circ$ 
from the plane due to the long tails of the PSF at low energies.

To determine the uncertainty of the
source and UIB components, we modified the effective 
area to the
extremes of its systematic uncertainty defined before and refitted the data.
Since the DGE model components do not vary in the fit, 
the absolute change in intensity caused by the modification to the effective
area propagates directly to the source and UIB components.
The systematic uncertainty on these components is energy dependent and due to
several effects.

For energies $\gtrsim 10$ GeV the PSF is $\sim 0.2^\circ$ (68\% containment) and
the sources are well-localised spatially.
Since the model is fixed and the sky maps are sparser at high latitudes for the 
data taking period in this paper, the UIB component 
absorbs almost
all of the intensity from the modification to the 
effective area.
At low energies the PSF is wider, $3.5^\circ$ (68\% containment) at 100 MeV for
\gray{} conversions in the front section of the LAT, and the 
sources are less well-localised spatially.
In addition, the sky maps are well populated even at high latitudes 
and display spatial structure.
The PSF broadening of the sources provides spatial structure and 
because the DGE model is fixed, more intensity is assigned to the source 
component to compensate in the fit.
These effects lead to the systematic error in the source component
being relatively larger than the isotropic at low energies and vice
versa at high energies.
Note, this applies for the high-latitude region from where the UIB 
component is derived, and also for the mid-latitude range for which we show
the combined contribution by sources in Fig.~\ref{fig:model}.
Because the uncertainties in the source and UIB components 
are not independent we have conservatively added their systematic uncertainties
for the total intensity band shown in Fig.~\ref{fig:model}.

The UIB component comprises the
true extragalactic diffuse \gray{} emission, 
emission from unresolved Galactic and extragalactic sources, and 
residual particle 
backgrounds (CRs that pass the \gray{} classification analysis 
and \gray{s} produced by CR
interactions in the passive material outside the ACD) in the LAT data.
In addition, other relevant 
foreground components that are not completely 
modelled, such as emission from the 
solar disk and extended emission \cite{Moskalenko2006} and
other potentially relevant 
``diffuse'' sources \cite{MP2007} are included.
Hence, the UIB component does {\em not} 
constitute a measurement of the extragalactic diffuse emission. 
Furthermore, comparison with the EGRET estimate of the extragalactic diffuse
emission \cite{sreekumar1998} is problematic due to the different DGE models
used and analysis details that are beyond the scope of the current paper and
will be addressed in a subsequent publication \cite{egbpaper}.

\begin{table}
\caption{\label{tab:table1}LAT data and components: 
$10^\circ \leq |b| \leq 20^\circ$.}
\begin{tabular}{ccccccccrrrrcc}
Energy \footnotemark[1] & 
LAT\footnotemark[2]\footnotemark[5] 
&& 
\multicolumn{4}{c}{Model\footnotemark[2]\footnotemark[3]\footnotemark[4]} 
&&
UIB\footnotemark[2]\footnotemark[5]\footnotemark[6] &
Source\footnotemark[2]\footnotemark[5] \\
\hline\hline
100--158 & $59.8 \pm 0.3$    && 26.0 & 11.0 & 6.4 & 8.6 && $21.0 \pm 0.1$ & $7.4 \pm 0.4$\\
158--251 & $65.0 \pm 0.3$    && 33.5 & 18.2 & 7.3 & 8.0 && $20.5 \pm 0.1$ & $5.4 \pm 0.1$\\
251--398 & $67.1 \pm 0.3$    && 38.2 & 23.2 & 7.6 & 7.4 && $18.7 \pm 0.1$ & $4.7 \pm 0.1$\\
398--631 & $64.5 \pm 0.3$    && 38.9 & 25.3 & 7.0 & 6.6 && $15.4 \pm 0.1$ & $3.8 \pm 0.1$\\
631--1000 & $60.8 \pm 0.3$   && 37.3 & 25.7 & 5.7 & 5.9 && $12.9 \pm 0.1$ & $3.2 \pm 0.1$\\
1000--1585 & $55.1 \pm 0.4$  && 32.8 & 23.3 & 4.4 & 5.1 && $11.6 \pm 0.1$ & $3.0 \pm 0.1$\\
1585--2512 & $46.3 \pm 0.4$  && 26.5 & 19.0 & 3.1 & 4.4 && $10.7 \pm 0.1$ & $2.7 \pm 0.1$\\
2512--3981 & $37.0 \pm 0.5$  && 20.2 & 14.4 & 2.0 & 3.8 && $9.2 \pm 0.1$ & $2.3 \pm 0.1$\\
3981--6310 & $29.9 \pm 0.5$  && 14.9 & 10.5 & 1.2 & 3.2 && $8.5 \pm 0.1$ & $1.9 \pm 0.1$ \\
6310--10000 & $20.7 \pm 0.5$ && 10.9 &\ 7.5 & 0.7 & 2.7 && $6.8 \pm 0.1$ & $1.5 \pm 0.1$ \\
\hline
\end{tabular}
\footnotetext[1]{MeV}
\footnotetext[2]{$E_\gamma ^2 J(E_\gamma)$ (10$^{-4}$ MeV cm$^{-2}$ s$^{-1}$ sr$^{-1}$) evaluated at the mid-bin energy.}
\footnotetext[3]{Total/$\pi^0$-decay/bremsstrahlung/inverse Compton.}
\footnotetext[4]{The GALPROP galdef ID for this model
is 54\_5gXvarh7S which is available at the website 
http://galprop.stanford.edu.}
\footnotetext[5]{Statistical errors only.} 
\footnotetext[6]{Unidentified background.}
\end{table}

{\it Discussion:}
The intensity scales of the LAT and EGRET have been found to be different with
the result that the LAT-measured spectra are softer.
In our early study of the Vela spectrum \cite{velapaper}, which was made 
using pre-launch IRFs, the difference was apparent already above 1 GeV.
Following on-orbit studies new IRFs have been developed to account for 
inefficiencies in the detection of \gray{s} in the LAT due to pile-up and 
accidental coincidence effects in the detector subsystems.
The inefficiency increases at lower energies, with the result that the IRFs 
used in the present analysis indicate greater intensities in the range below 
1 GeV, with the magnitude of the effect ranging up to $\sim$30\% at 100 MeV.
A forthcoming study of the Vela pulsar using the LAT one-year data with 
post-launch
IRFs also shows a similar effect in the low-energy pulsed spectrum. 
So, the relative brightness of the diffuse emission measured by the
LAT at low energies is unlikely to be due to increased residual background.
Our confidence that the IRFs used in the present analysis accurately represent
our knowledge of the instrument comes from detailed instrument simulations that 
were validated with beam tests of calibration units, and to post-launch 
refinements based on actual particle backgrounds.
The systematic uncertainty on the effective area gives an energy dependent
measure of our confidence in the IRFs used in the present analysis.

As a consequence, 
the LAT-measured DGE spectrum averaged over all Galactic longitudes for the 
latitude range $10^\circ \leq |b| \leq 20^\circ$ is systematically softer 
than the EGRET-measured spectrum.
The spectral shape is compatible with that of an {\em a priori} DGE model that 
is consistent with directly measured CR spectra.
The excess emission above 1 GeV measured by EGRET is not seen by the LAT 
in this region of the sky.

While the LAT spectral shape is consistent with the DGE model used in this 
paper, the overall model emission is too low thus giving rise to a 
$\sim10-15$\% excess over the energy range 100 MeV to 10 GeV.
However, the DGE model is based on pre-\fermi{} data and knowledge
of the DGE.
The difference between the model and data is of the same order as the 
uncertainty in the measured CR nuclei spectra at the relevant 
energies \cite{besspaper}.  
In addition, other model parameters that can affect the \gray{} production 
rate (e.g., the conversion between CO line intensity and molecular
hydrogen column density in the interstellar medium, \Xco) have not been 
modified in the present paper.
Overall, 
the agreement between the LAT-measured spectrum and the model shows that 
the fundamental processes are consistent with our data, thus providing a
solid basis for future work understanding the DGE.

{\it Acknowledgements:}
The \fermi{} LAT Collaboration acknowledges support from a number of 
agencies and institutes for both development and the operation of the LAT 
as well as scientific data analysis. 
These include NASA and DOE in the United States, CEA/Irfu and IN2P3/CNRS 
in France, ASI and INFN in Italy, MEXT, KEK, and JAXA in Japan, and 
the K.~A.~Wallenberg Foundation, the Swedish Research Council and the 
National Space Board in Sweden. 
Additional support from INAF in Italy for science analysis during the 
operations phase is also gratefully acknowledged.

GALPROP development is partially funded via NASA grant NNX09AC15G.

Some of the results in this paper have been derived using the HEALPix \cite{healpix} package.


\bibliography{ms_arxiv}

\end{document}